\documentclass[final]{IEEEtran}

\usepackage{amsthm,amssymb,graphicx,multirow,amsmath,color,amsfonts}
\usepackage[update,prepend]{epstopdf}
\usepackage[noadjust]{cite}
\usepackage[latin1]{inputenc}
\usepackage{tikz}
\usetikzlibrary{arrows,calc}		
\usepackage{bbm} 
\usepackage{pdfpages}
\usepackage{booktabs}
\usepackage{tabulary}
\usepackage{multirow}
\usepackage{comment}
\usepackage{textcomp}
\usepackage[bottom]{footmisc}

\include{notation}
\allowdisplaybreaks 

\newcommand*\circled[1]{\tikz[baseline=(char.base)]{
            \node[shape=circle,draw,inner sep=1pt] (char) {#1};}}
\renewenvironment{IEEEbiography}[1]
  {\IEEEbiographynophoto{#1}}
  {\endIEEEbiographynophoto}

\begin{document}
\title{Intelligent O-RAN for Beyond 5G and 6G Wireless Networks}

\author{
Solmaz Niknam, Abhishek Roy, Harpreet S. Dhillon, Sukhdeep Singh, Rahul Banerji, Jeffery H. Reed, Navrati Saxena, and Seungil Yoon
\thanks{S. Niknam, H. S. Dhillon, and J. H. Reed are with Wireless@VT, Department of Electrical and Computer Engineering, Virginia Tech, Blacksburg, VA ({\texttt {email: \{slmzniknam, hdhillon, reedjh\}@vt.edu}}).
A. Roy is with MediaTek, San Jose, CA ({\texttt {email: abhishek.roy@mediatek.com}}). This work was done when he was with Samsung Electronics, Suwon, South Korea. S. Singh, and R. Banerji are with Samsung R\&D, Bangalore, India ({\texttt {email: \{sukh.sandhu, r.banerji\}@samsung.com}}). N. Saxena is with the Department of Computer Science and Engineering, Sungkyunkwan University, South Korea ({\texttt {email: navrati@skku.edu}}). S. Yoon is with Samsung Electronics, Suwon, South Korea ({\texttt {email: siyoon72@samsung.com}}). \newline
This work is supported in part by the U.S. NSF (Grants CNS-1564148, CNS-1814477), Oak Ridge National Laboratory (Grant 4000170832), and Commonwealth Cyber Initiative. 
}
}
\maketitle

\begin{abstract}
Building on the principles of {\em openness} and {\em intelligence}, there has been a concerted global effort from the operators towards enhancing the radio access network (RAN) architecture. The objective is to build an operator-defined RAN architecture (and associated interfaces) on open hardware that provides intelligent radio control for beyond fifth generation (5G) as well as future sixth generation (6G) wireless networks. Specifically, the open-radio access network (O-RAN) alliance has been formed by merging xRAN forum and C-RAN alliance to formally define the requirements that would help achieve this objective. Owing to the importance of O-RAN in the current wireless landscape, this article provides an introduction to the concepts, principles, and requirements of the Open RAN as specified by the O-RAN alliance. In order to illustrate the role of intelligence in O-RAN, we propose an intelligent radio resource management scheme to handle traffic congestion and demonstrate its efficacy on a real-world dataset obtained from a large operator. A high-level architecture of this deployment scenario that is compliant with the O-RAN requirements is also discussed. The article concludes with key technical challenges and open problems for future research and development.
\end{abstract}

\section{Introduction} \label{sec:intro}
The fifth generation (5G) cellular network has been standardized to meet diverse demands that are classified into three broad categories, including enhanced mobile broadband (eMBB), ultra-reliable and low-latency communications (uRLLC), and massive machine type communications (mMTC).
However, the existing 5G wireless architecture lacks sufficient flexibility and intelligence to efficiently handle these demands~\cite{Roadmap6GLetaief,6G2019Mingh}. As a result, the evolution towards beyond 5G and sixth generation (6G) wireless calls for an architectural transformation required to support service heterogeneity, coordination of multi-connectivity technologies, and on-demand service deployment.
Open radio access network (RAN) is an emerging idea that enables such a transformation using the concepts of virtualization, flexibility, and intelligence. Naturally, over the last few years, multiple independent alliances and forums have initiated research on accelerating this transformation of RAN by increasing infrastructure virtualization, combined with embedded intelligence to deliver more agile services and advanced capabilities to end users.

One such effort is known as OpenRAN, a project group within the Telecom Infra Project (TIP), that focuses on building RAN solution based on software-defined technology and open and general-purpose hardware~\cite{Roy2019accenture}.
Another separate effort is xRAN forum that has been formed to promote an open alternative to traditionally vendor-based RAN architecture. The xRAN effort is focused towards advancing RAN in three areas including separation of user and control planes, open interfaces, and modular RAN software stack on commercial off-the-shelf (COTS) hardware~\cite{xran2016forum}.
On February 2018, open-radio access network (O-RAN) was conceived by merging xRAN forum and C-RAN alliance to drive new levels of openness in the radio access network that would support the evolution towards beyond 5G and 6G wireless.
The main objective of O-RAN is to enhance the RAN performance through virtualized network elements and open interfaces that incorporate intelligence in RAN. \emph{Openness} and \emph{intelligence} are the two core pillars of the efforts pursued by the O-RAN alliance, which is a global force consisting of more than 160 contributors from large vendors, small and medium companies, network operators, start-ups and academic institutions~\cite{alliance2018ran}.
Openness aims to eliminate vendor lock-in and proprietary implementation of hardware and software by establishing
open standard RF interfaces, which help in increasing operational savings already provided by virtual RAN (vRAN) and cloud RAN (C-RAN). This will enable the deployment of remote radio heads (RRHs) and baseband units (BBUs) from different vendors to build flexible and scalable RAN networks. Besides flexibility, openness of RAN components accelerates the delivery of new features and services where services can be dynamically introduced to users.

Intelligence is quickly becoming a necessity for the deployment, optimization, and operation of wireless networks beyond 5G~\cite{AI2020Renzo,niknam2019federated}. This is primarily because of the increasing complexity of 5G wireless networks and beyond, in response to the need to handle demanding service requirements~\cite{challita2019machine,ML2018simone}. Therefore,
``\emph{O-RAN alliance strives to leverage emerging learning techniques to embed intelligence in every layer of the RAN architecture. Embedded intelligence, applied at both component and network levels, enables dynamic local radio resource allocation and optimizes network-wide efficiency}"~\cite{alliance2018ran}. Ratification of several specifications and release of millions of lines of open-source code (in partnership with Linux) is gradually establishing O-RAN as the harbinger of a collaborative platform to support the evolution towards the next generation of wireless communication networks.

Owing to the implications of these developments on the future wireless and networking research, it is imperative to provide a timely and accessible introduction to the general concept and core principles of O-RAN, so that these concepts can benefit from inputs from the broader community (and not just the current stakeholders, which are mostly the operators). Keeping this rather ambitious goal in mind, we take a two-pronged approach in this article. We first provide a brief introduction to O-RAN to educate readers about the general concept, while providing pointers for more advanced reading (which is necessary because of the space constraints). We then focus specifically on illustrating the role of intelligence in O-RAN, which we believe will be an essential factor moving forward. For this, we propose an intelligent traffic prediction and radio resource management scheme that is cognizant of the O-RAN architectural requirements. This scheme is described next.

We utilize long short-term memory (LSTM) recurrent neural network (RNN) to learn and predict the traffic pattern of a real-world cellular network in a densely populated area of Mumbai, India, in order to identify potential congested cells. The LSTM model is trained at non-real-time radio intelligence controller (non-RT RIC) in the O-RAN architecture, using long term data gathered from RAN. The trained model is then sent to near-real-time radio intelligence controller (near-RT RIC) of the O-RAN for inference. Upon the inference outcome, cell splitting is applied to the congested cells to improve the related key performance indicators (KPIs). Traffic prediction and the corresponding congestion treatments are continuously applied until the target KPI values are met. In order to show the compliance of the overall scheme with O-RAN requirements, we also discuss how the proposed mechanism is mapped into the O-RAN control loops, specify the location of the machine learning (ML) training and inference modules, and provide a high-level architecture of deployment scenarios and the end-to-end flow. 
{To the best of our knowledge, this article makes
the first attempt to demonstrate a concrete O-RAN based practical example with embedded intelligence.}

\begin{figure}[t]
\centerline{\includegraphics[scale=0.3]{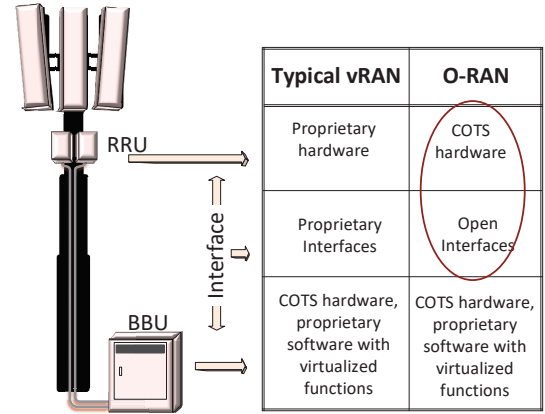}}
\caption{Comparison of O-RAN and vRAN approaches in terms of openness. The major difference is in RRU hardware and the RRU-BBU interface.}
\label{fig:ORAN_vRAN}
\centering
\end{figure}

\section{Preliminaries And Overview}  \label{sec:Prelim}
The cost involved in the deployment, optimization and operation of the RAN components generally accounts for approximately 70\% of the total network cost~\cite{Roy2019accenture}.
This is one of the main reasons behind considering RAN as the most appealing candidate by operators for decreasing the network expenditure. One of the primary RAN architectures introduced to enable cost saving on expensive baseband resources has been the C-RAN architecture, in which the baseband units are shared in a centralized baseband pool. Therefore, the computing resources can be utilized optimally based on the demand. This architecture has opened up an opportunity for RAN virtualization that further reduces cost. As a result, vRAN has been developed to simplify the deployment and management of the RAN nodes and make the platform readily available for multitude of dynamically changing service requirements.

Although quite cost-effective, these architectures still host propriety software, hardware and interfaces. In fact, lack of openness has been identified as a major bottleneck in maximally utilizing virtualization~\cite{sam2019challenge}.
Please refer to Fig.~\ref{fig:ORAN_vRAN} for the vRAN architecture.

\begin{figure}[t]
\centerline{\includegraphics[scale=0.47]{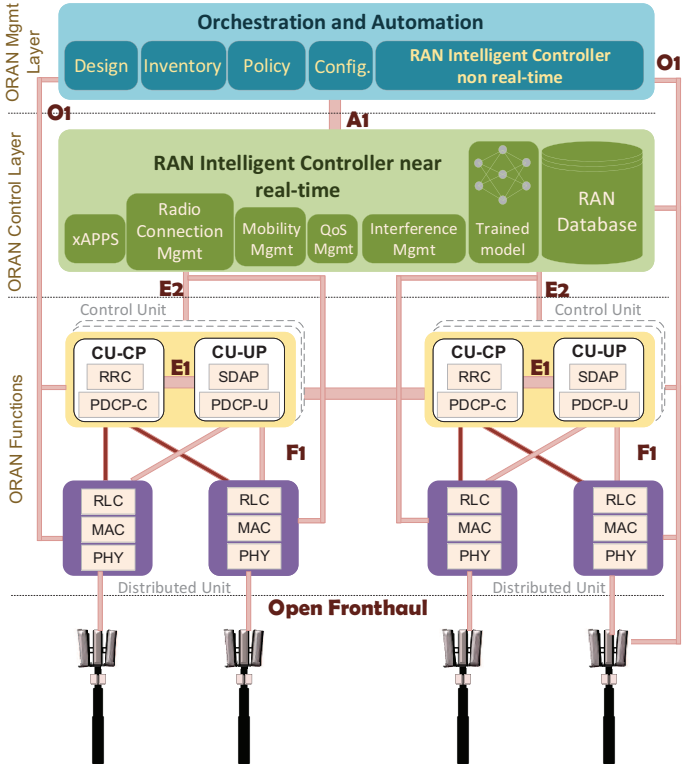}}
\caption{Representation of O-RAN architecture~\cite{alliance2018ran}, with RAN intelligent controllers (near real-time and non real-time), control and distributed units.}
\label{fig:ORAN_Arch}
\centering
\end{figure}

In order to overcome the limitations of C-RAN and vRAN, O-RAN is emerging as a new RAN architecture that uses well-defined open interfaces between the elements implemented on general-purpose hardware. It also allows RRU and BBU hardware and software from different vendors (see Fig.~\ref{fig:ORAN_vRAN}). Disaggregation is a key factor based on which operators can select RAN components from different vendors individually. In addition, open interfaces between decoupled RAN components provide efficient multi-vendor interoperability.
Another major tenet of O-RAN architecture is RAN virtualization. Enhancing virtualization supports more efficient splits over the protocol stack for network slicing purpose.
To further reduce the RAN expenditure, O-RAN fosters self-organizing networks, that reduces conventional labor intensive means of network deployment, operation and optimization.
In addition to cost reduction, intelligent RAN can handle the growing network complexity and improve the efficiency and accuracy by reducing the human-machine interaction. Radio intelligent controllers, non-RT RIC and near-RT RIC, are two main modules introduced in O-RAN architecture that enhance the traditional network functions with embedded intelligence (see Fig.~\ref{fig:ORAN_Arch}).
The near-RT RIC is further interfaced with centralized unit control plane (CU-CP) and centralized unit user plane (CU-UP), which are responsible for signaling and configuration messages, and data transmission, respectively. Distributed unit (DU) access to centralized units and provide services for users through RRHs.
O-RAN alliance strives to steer the industry towards the development of artificial intelligence (AI)-enabled RICs~\cite{from2020kovski}.
There are several key steps that need to be taken in any ML/AI-assisted solution, based on the O-RAN requirements~\cite{ORAN2019WG2}.
\begin{itemize}
\item The first step is \emph{model capability query} that is performed by the service management and orchestration (SMO), when the model is to be executed for the first time (or updated). These capabilities include hardware processing power, ML engine, and available data sources.

\item The next step is \emph{model selection and training}, where the ML training host initiates the model training and sends the trained model back to the non-RT RIC in SMO for deployment.

\item The ML inference host is then configured with the model description file, and the online data shall be used for inference. The inference outcome is sent to near-RT RIC, from where the policy is generated to take corrective actions.

\item Depending on the outcome of the model inference, the corresponding actions are taken using the related actors. Based on the location of the ML inference and the actors and type of actions, different interfaces (O1, A1 and E2) are utilized.
{A1 interface is an open logical interface to enable the non-RT RIC to provide policy-based guidance, ML model management, and enrichment information to the near-RT RIC function for RAN optimization.}

{E2 is the interface between near-RT RIC, the centralized unit (CU) protocol stack and the underlying RAN DU. This provides a standard interface between the near-RT RIC and CU/DU in the context of O-RAN architecture.}

{The role of O1 interface is to provide operation and management of CU, DU, radio unit (RU), and near-RT RIC (such as fault management, performance management and configuration management) to SMO. It can also configure CU, DU, RU, and near-RT RIC depending on the use cases~\cite{Use2019Oran} which is beyond the scope of this article.}

\item Finally, upon \emph{monitoring the performance of the model}, the inference host feeds back the model performance to the training host for the purpose of \emph{model redeployment or model update}.
\end{itemize}
There are some initial set of exemplary use cases (showcasing the utilization of ML/AI models), including context-based dynamic handover management for vehicle-to-everything (V2X) communication, quality of experience (QoE) optimization, and flight path-based dynamic unmanned aerial vehicle (UAV) resource allocation, to demonstrate the practical applicability of O-RAN architecture. Interested readers are referred to~\cite{Use2019Oran} for additional use cases.
\section{Proposed Framework}  \label{sec:Proposed}
In order to provide a concrete practical example, we develop an intelligent radio resource management scheme tailored for the O-RAN architecture. In this scheme, the spatial pattern of the data traffic is learned by utilizing LSTM neural network to predict the possible occurrence of the congestion. To prevent an upcoming congestion, radio resources are re-allocated accordingly.
In fact, the objective is to utilize a COTS learning technique and embed intelligence in O-RAN architecture.
This section provides detailed description of the proposed scheme and its compatibility with O-RAN requirements on the ML/AI-assisted solutions described in~\cite{ORAN2019WG2}.
\begin{figure*}[t]
\centerline{\includegraphics[scale=0.25]{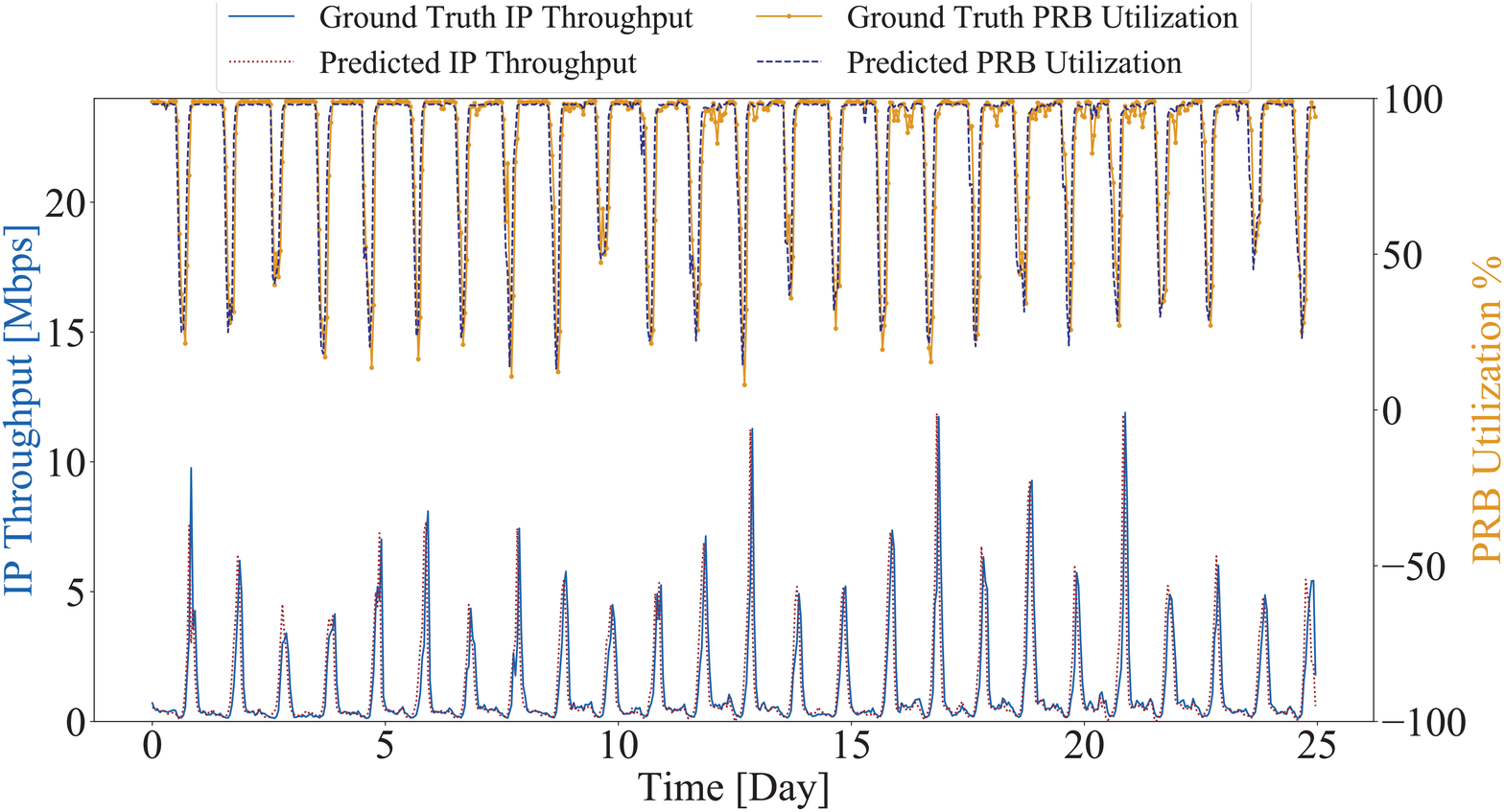}}
\centering
\caption{User-perceived IP throughput and PRB utilization prediction for a cell of a selected eNB in the network. }
\label{fig:predict}
\end{figure*}
\begin{figure}[t]
\vspace{0.35cm}
\centerline{\includegraphics[scale=0.19]{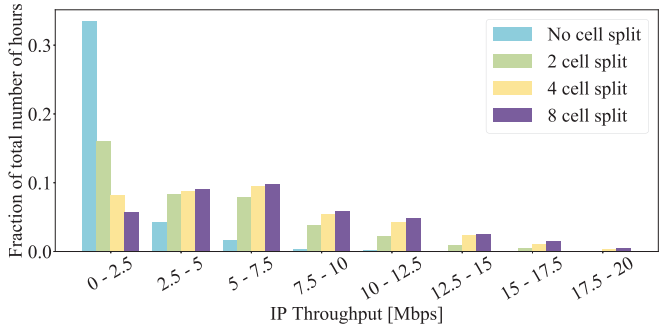}}
\caption{User-perceived IP throughput performance for different order of cell splitting.}
\label{fig:IP_throu}
\centering
\end{figure}
\subsection{Intelligent Radio Resource Management}
In the proposed scheme, we define a cell as \emph{congested} if
\begin{itemize}
\item The average user-perceived IP throughput $<$ 1Mbps; \textbf{AND}
\item The average downlink physical resource block (DL-PRB) utilization $>$ 80\%;  
\end{itemize}
DL-PRB utilization percentage provides the usage (in percentage) of PRBs on the downlink for user plane traffic. In addition,
the user-perceived IP throughput is measured in terms of the packets transmitted between the evolved node-B (eNB) and users~\cite{ETSIts132}.
It is worth mentioning that the above metrics to identify the congestion event and the corresponding thresholds are defined based on the operator service level agreement (SLA) and can be re-configured by the operator based on their hardware or software requirements.
Following these metrics, network parameters such as
PRB utilization rate and {user\textquotesingle}s downlink data rate are continuously monitored across all cells of the eNBs in the network.
Using RNN, the temporal pattern of the mentioned parameters are learned through the current values to predict future values and the potential congested cells.
Subsequently, the network alarms are set to trigger if the cells are likely to get congested. It is worth noting that different triggering criteria could be considered based on the target KPIs. Upon identifying the congested cells, solutions, such as enabling dual connectivity, and cell splitting can be applied as remedies. Finally, if the prediction is erroneous, the weights of the RNN model are updated based on the actual value of the parameter to reflect the changes and improve the performance until the target KPI conditions are met.

The parameters of an RNN model that include 2 layers of 12 LSTM units, are learned to predict the future traffic for the next hour. This can be configured by operators as per the available data and its periodicity. The RNN training is carried out over a real-world mobile traffic dataset from a cellular network in Mumbai, India. The dataset contains network measurements in terms of user-perceived IP throughput, downlink PRB utilization, collected from 17 LTE eNBs (18 cells in each eNB), over a duration of 25 days, August 1 to August 25, 2019. Simulation parameters are summarized in Table~\ref{tab:sim_param}.
The RNN model is implemented using Keras, the open-source high-level TensorFlow application programming interface. The model training is carried out on a server with dual Xeon Gold CPU (44 threads/CPU) along with 512~GB RAM.
In order to illustrate the performance of the ML model prediction, Fig.~\ref{fig:predict} represents the performance of the RNN model in terms of user-perceived IP throughput and percentage of DL-PRB utilization. In this figure, both actual and predicted values for user-perceived IP throughput (left y-axis) and percentage of DL-PRB utilization (right y-axis) of a cell in a selected eNB in the network are shown. The average accuracy of the prediction is 92.64\%.
\begin{table}[t]
\caption{Simulation Parameters}
\centering
\begin{tabular}{l|c}
\multicolumn{1}{c}{\textbf{Parameter}}  & \textbf{Value} \\
\hline
No. of eNB  & 17\\
No. of cell in each eNB & 18\\
$\%$ of the cell splitting ($R$)& \multicolumn{1}{l}{{[}60,75{]}} \\
LSTM layer   & 2\\
No. of LSTM unit in each layer & 12  \\
Batch size  & 16 \\
No. of epoch & 150\\
Activation function & $\tanh$ \\
Optimizer & Adam\\
\hline
\end{tabular}\label{tab:sim_param}
\end{table}

\begin{figure*}[t]
\centering
\centerline{\includegraphics[scale=0.5]{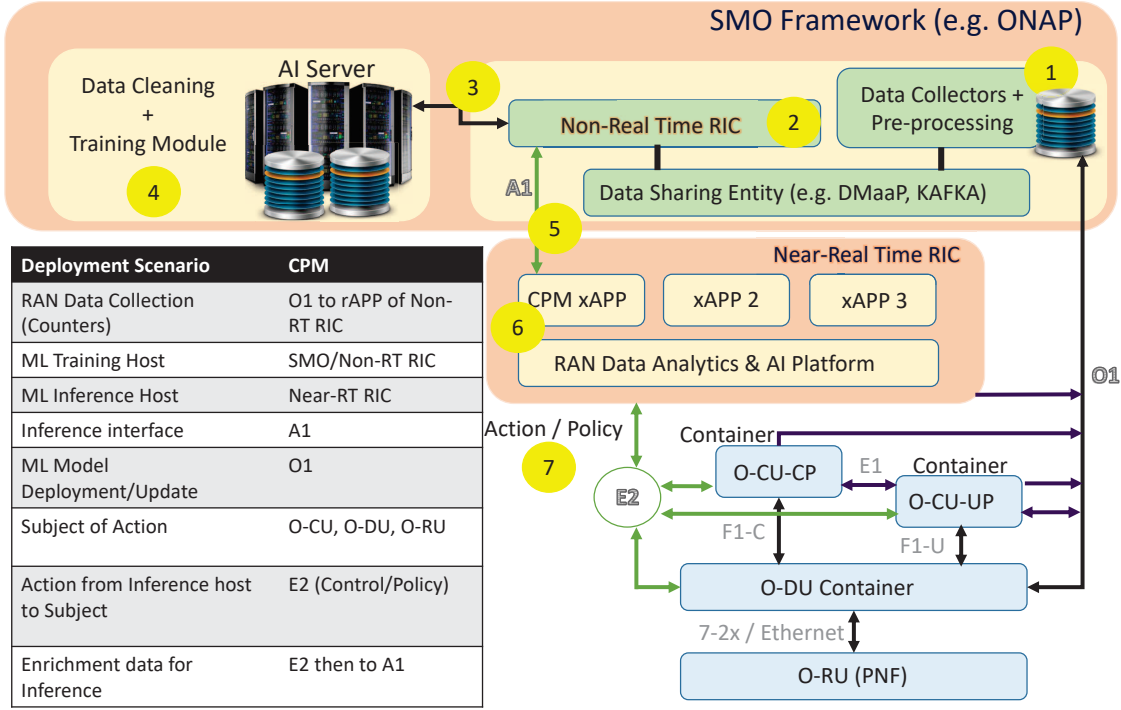}}
\caption{High-level structure of deploying the proposed intelligent congestion prediction and radio resource management scheme in the O-RAN architecture.}
\label{fig:TRIO_Archit}
\centering
\end{figure*}
As a congestion relief solution, we utilize cell splitting approach. Here, cell splitting refers to the general idea of splitting the coverage footprint of congested cells into two or more cell sites. For the purpose of this discussion, one can achieve cell splitting by activating eNBs (especially, small cells) that may be put into sleep mode during highly loaded periods to save energy. Since power consumption is one of the biggest costs for the operators, strategies that are ``green'' are of strategic importance to them. Another economically viable way of doing this is through infrastructure sharing. For instance, instead of operators building their own systems to accommodate peak traffic activity, they can rely on infrastructure sharing to access additional cells or offload their additional traffic to base stations operating in unlicensed spectrum, such as citizens broadband radio service (CBRS). This strategy complements emerging business models of both network and spectrum sharing to increase network capacity.

That said, since our focus is on demonstrating how intelligence can be embedded in O-RAN, the exact choice of the congestion solution is immaterial; because of which we selected a simple scheme that is easy to describe on a real network.
As a result of cell splitting, a fraction of users in the original cell are moved to the splitted cells.
Therefore, in order to emulate such an effect in our simulation, a random number $R \in [60,\,75]$ is generated in each round of split. Subsequently, $R\%$ of the users are assumed to move to the new cell, while the remaining users stay in the original cell.
For higher cell splitting factors, i.e. 4 or 8, the same process is repeated for the splitted cells in each round. Fig.~\ref{fig:IP_throu} demonstrates the network performance in terms of user-perceived IP throughput. Each bar in the histogram represents the number of hours that the user-perceived IP throughput of the given cell is within a certain interval, specified by the range of the bar on the IP throughput-axis. As evident from the result, the preemptive cell splitting of the congested cells in the network significantly improves the performance. Although, as seen in the figure, one can achieve a higher network capacity by more aggressive cell splitting (due to denser frequency reuse), it can stress some other factors, such as increasing the complexity of channel assignment and increasing the occurrence of handovers, to name a few. Therefore, cell splitting should be implemented in moderation to ensure that the cell congestion is avoided with minimal degradation in the aforementioned factors.
\subsection{Deployment Architecture}
In this section, we explain how the proposed scheme is implemented in the O-RAN architecture.
The high-level structure of deployment scenarios and end-to-end flow of the proposed solution in the O-RAN architecture are illustrated in Fig.~\ref{fig:TRIO_Archit}.
\begin{itemize}
\item[\circled{1}] The related RAN counters from control  and distributed units are collected in the data collector located in the SMO. Depending on the SMO platform, different entity would be responsible for data collection~\cite{ORAN2019WG2}. For instance, if open network automation platform (ONAP) is considered as the SMO, virtual event streaming (VES) collectors of data collection, analytics and events (DCAE) subsystem in the ONAP is used to collect the data. This step is carried out over O1 interface of the O-RAN architecture. It is worth noting that assuming DCAE, some data preprocessing such as adding virtual network function (VNF) names and IDs and converting counters into KPIs are carried out by open-source cask data application platform (CDAP).
\item[\circled{2}] The collected data at the SMO is shared with non-RT RIC deployed in the SMO using a data bus, such as Kafka.
\item[\circled{3}] The related ML/AI model, hosted in the AI server inside the SMO, is queried by non-RT RIC. We have utilized Acumos AI to deploy the training module. One can of course build other AI platforms inside DCAE.
\item[\circled{4}] After training the model in the AI server, the inference is sent back to non-RT RIC.
\item[\circled{5}] Subsequently, the inference results and policies are forwarded to congestion prediction and mitigation (CPM) xAPP in near-RT RIC over the A1 interface of the O-RAN. xAPPs are applications that are specific to radio-function to make the RAN components programmable.
Fig.~\ref{fig:TRIO_Archit} provides details of the ML training and inference host locations (non-RT RIC and near-RT RIC, respectively) in our proposed architecture. This is based on the second set of deployment scenarios\footnote{Depending on the training and inference locations, there are three different deployment scenarios specified by O-RAN alliance. Interested readers are referred to~\cite{ORAN2019WG2}.} specified by the O-RAN alliance in technical report~\cite{ORAN2019WG2}.
\item[\circled{6}] The congestion relief solution is configured once the congestion occurrence is predicted.
\item[\circled{7}] Finally, the corresponding solution is applied to CU or DU through E2 interface.
\end{itemize}

\section{Challenges and Open Problems} \label{sec:challenge}
In an actual wireless network, a group of cells may have specific performance patterns and infrastructural requirements that would depend on their locations within the network and subsequently the statistics of the load they are serving. Information about these specific requirements will help to increase the efficacy of models for self-estimation and self-healing of congestion. This includes analysis of busy-hour traffic patterns and the associated atmospheric conditions across ultra-dense cells in major metropolitan cities. Typically with increasing traffic requirements, as the PRB usage exceeds a certain threshold, the wireless network operator either adds more cells or introduces additional carriers. Our system should be capable enough to create a model that is adaptable for such dynamic cell and carrier additions. Interestingly, as O-RAN introduces the concept of openness, an operator can have different equipment from different vendors. All the operators and vendors have different naming conventions for counters and KPIs. Thus, the model needs to be flexible enough to dynamically adapt to different metadata from different operators and vendors. Moreover,  difference in hardware and software  performance often results in operators and vendors having different SLAs to measure cell congestion. In such scenarios, we expect the vendors and operators to either cooperate in agreeing on common SLAs or improve our models to dynamically adapt based on different SLAs. Keeping inference models for thousands of cells in near-RT RIC is quite complex. Such complex models are generally not efficient for execution in existing high performance CPUs and might require GPUs to take care of this complexity, which in turn, might involve higher capital and operating expenditures.

Since O-RAN stems from the key principle of RAN virtualization, it inherits deployment-specific security challenges attributed by virtualization and software defined network (SDN). These security challenges include authentication and authorization of virtual machine (VM) migration, VM instantiation,
hypervisor security, orchestration security, and SDN controller security~\cite{firoozjaei2017security}.
In addition, the shared BBU pool in the O-RAN cloud native deployment may impose the risk of breaking user privacy and accessing sensitive data.
Therefore, although O-RAN enables the creation of flexible service tailored to the needs of distinct customers, it is important to weigh these benefits in the light of security challenges brought in by the open and virtual approaches.

Extreme data rate requirement of eMBB application stretches the limits of common public radio interface (CPRI)-based fronthaul. In addition, large bandwidth requirement in CPRI fronthaul limits the cloud native deployment, which is an integral part of the O-RAN vision. However, utilizing an Ethernet-based transport through the 7.2x specification in the O-RAN architecture has moderated this limitation~\cite{Use2019Oran}. While Ethernet has been able to meet stringent data rate requirement, fronthaul transport requirements are significantly more challenging in network slices with uRLLC requirement, such as tactile Internet, industrial control and automotive applications. In such applications,  the network transport capacity may not be sufficient and using virtual BBU may add higher latency~\cite{Roy2019accenture}. Therefore, backup strategies for distinct use case scenarios should be in place.

Another challenge associated with open architectures that incorporate multi-vendor elements is interoperability. In fact, to maintain the stability and reliability of the operation in O-RAN, multi-vendor products must interoperate.
Furthermore, risk mitigation strategies should be in place, in case implementations do not work with each other successfully. Therefore, it is crucial to identify the risks of incompatibilities between the radio and control products from different vendors~\cite{sam2019challenge}.

\section{Summary and Concluding Remarks} \label{sec:Conclusion}
The O-RAN alliance is a world-wide effort conceived by merging xRAN forum and C-RAN alliance to drive new levels of openness and intelligence in the radio access network of next generation wireless systems. In this article, we started off by providing an accessible introduction to the general concept of O-RAN and its two core principles of openness and intelligence. In order to provide a concrete O-RAN based practical example, the temporal pattern of a real-world data traffic from a dense urban cellular network in Mumbai, India, was learned by utilizing LSTM neural network to predict the possible congestion with high accuracy. In order to prevent an upcoming congestion, we discussed a cell-splitting based radio resource management scheme along with its corresponding high-level architecture (as well as the end-to-end flow) that is cognizant of the O-RAN requirements. Since our real objective in this article was to discuss architectural subtleties of embedding intelligence in O-RAN, we limited our attention to a specific congestion solution that is easy to describe in an actual network. We conclude this discussion with the hope that this article will convey the essence of O-RAN to the broader community to actively engage them in this exciting new area, which clearly has important implications for future communications and network research. In order to help the uninitiated, we have also provided pointers to several open research questions.

\bibliographystyle{IEEEtran}
\bibliography{IEEEabrv,GBbibfile}

\vskip 0pt plus -1fil
\begin{IEEEbiography}{Solmaz Niknam}
received her B.Sc. degree (1st class honor) in Electrical Engineering from Shiraz University of Technology, Shiraz, Iran, in 2010, her M.Sc. degree in Electrical Engineering from Iran University of Science and Technology, Tehran, Iran, in 2012 and her Ph.D. degree from Kansas State University, KS, USA in 2018. During her Ph.D., she was a recipient of the Kansas Ph.D. students Fellowship. She is currently a postdoctoral associate at Virginia Tech. Her research interests include wireless communication with emphasis on 5G mm-wave networks and ML/AI in communication.
\end{IEEEbiography}
\vskip 0pt plus -1fil
\begin{IEEEbiography}{Abhishek Roy}
is currently working as a senior technical manager at MediaTek. He received his Ph.D. degree in 2010 from Sungkyunkwan University, his M.S. degree in 2002 from the University of Texas at Arlington, and his B.E. degree in 2000 from Jadavpur University, India. He has strong professional skills in 4G/5G/6G wireless system design, New Radio unlicensed, IoT, cloud RAN, network modeling, and simulation.
\end{IEEEbiography}
\vskip 0pt plus -1fil
\begin{IEEEbiography}{Harpreet S. Dhillon} is an Associate Professor of Electrical and Computer Engineering and the Elizabeth and James E. Turner Jr. '56 Faculty Fellow at Virginia Tech. He received his B.Tech. degree from IIT Guwahati in 2008, his M.S. degree from Virginia Tech in 2010, and his Ph.D. degree from the University of Texas at Austin in 2013, all in Electrical Engineering. His research interests include communication theory, wireless networks, stochastic geometry, and machine learning. He is a Clarivate Analytics Highly Cited Researcher and a recipient of five best paper awards. He serves as an Editor for three IEEE journals.
\end{IEEEbiography}
\vskip 0pt plus -1fil
\begin{IEEEbiography}{Sukhdeep Singh}
is currently working as Chief Engineer (Technical Manager) at Samsung R\&D Bangalore, India. He received his Ph.D. from Sungkyunkwan University, South Korea in 2016. His research interest includes 4G/5G RAN system design, O-RAN, vRAN, Cloud Native and TCP/QUIC for 5G/6G cellular networks.
\end{IEEEbiography}
\vskip 0pt plus -1fil
\begin{IEEEbiography}{Rahul Banerji}
is currently working as Senior Software Engineer at Samsung R\&D India Bangalore. He received his B.E. from BITS Pilani, India in 2018. His research interest includes ML/AI modeling and simulations for Next Generation Mobile Networks, Software Design of vRAN and O-RAN systems, Service Management and Operation for 5G wireless.
\end{IEEEbiography}
\vskip 0pt plus -1fil
\begin{IEEEbiography}{Jeffrey H. Reed} is currently the Willis G. Worcester Professor with the Bradley Department of Electrical and Computer Engineering, Virginia Tech. He received his B.S., M.S., and Ph.D. degrees from the University of California, Davis, CA, USA, in 1979, 1980 and 1987, respectively. He is the founder of wireless@Virginia Tech and the Founding Faculty Member of the Ted and Karyn Hume Center for National Security and Technology. He is a Fellow of the IEEE for his contributions to software radio. He is the author of three books and over 300 journal and conference papers.
\vskip 0pt plus -1fil
\begin{IEEEbiography}{Navrati Saxena}
is an associate professor in the Department of Computer Science and Engineering, Sungkyunkwan University, South Korea. She completed her Ph.D. in the Department of Information and Telecommunication, University of Trento, Italy. Her prime research interests involve 4G/5G wireless, IoT, smart grids, and smart environments.
\end{IEEEbiography}
\vskip 0pt plus -1fil
\begin{IEEEbiography}{Seungil Yoon}
is principle engineer at Samsung Electronics, Head Quarters, South Korea. He has worked with Samsung for 16 years. He completed his Ph.D. from Georgia Institute of Technology in 2011. His research interests include mobility for network slicing, micro-services based network architecture, O-RAN and vRAN.
\end{IEEEbiography}

\end{IEEEbiography}

\end{document}